\documentclass[a4paper]{jpconf}
\usepackage{graphicx}

\begin{document}

\title{Searching for white dwarfs candidates in Sloan Digital Sky Survey Data}

\author{Miros{\l}aw Nale{\.z}yty$^{1}$,
        Agnieszka Majczyna$^{1,2}$,
        Anna Ciechanowska$^{1}$ and
        Jerzy Madej$^{1}$}
\address{$^1$University of Warsaw Astronomical Observatory,
         Al. Ujazdowskie 4,
         00-478 Warsaw,
         Poland}
\address{$^2$The Andrzej So{\l}tan Institute for Nuclear Studies,
         Ho{\.z}a 69,
         00-681 Warsaw,
	 Poland}

\ead{nalezyty@astrouw.edu.pl}

\begin{abstract}
Large amount of observational spectroscopic data are recently
available from different observational projects, like Sloan Digital
Sky Survey. It's become more urgent to identify white dwarfs stars
based on data itself i.e. without modelling white dwarf
atmospheres. In particular, existing methods of white dwarfs
identification presented in Kleinman et al. (2004) and in Eisenstein
et al. (2006) did not allow to find all the white dwarfs in examined
data. We intend to test various criteria of searching for white dwarf
candidates, based on photometric and spectral features.
\end{abstract}

\section{Introduction}

Because of their physical properties, white dwarfs stars are difficult
to observe and identification. The first version of the McCook \& Sion
catalogue (1987) contains 1279 white dwarfs only. Recently,
observational techniques was considerable improve, so the amount of
photometric and spectral data rapidly grows. For this reason also the
number of identified white dwarfs grows, but it is very needed to
elaborate efficient identification and classification methods.

Until now, many of the white dwarf samples were selected, based on
different selection criteria. For example Fleming et al. (1996)
published a white dwarfs catalogue based on ROSAT data (Trumper 1983,
1992). A classification was made by searching for position coincidence
between ROSAT sources and hot white dwarfs from McCook \& Sion
catalogue. Sources with no optical counterparts were marked as white
dwarf if was visible only in the softest X-ray energy range. White
dwarfs in EUV, selected also from the ROSAT all-sky survey by Marsh et
al. (1997a,b), 129 white dwarf stars from Bergeron, Saffer \& Liebert
(1992), 200 white dwarfs from the Palomar Green survey (Green, Schmidt
\& Liebert (1986) analyzed by Liebert \& Bergeron (1995) - it is only
a few samples of these objects.

The Sloan Digital Sky Survey project (York et al. 2000) provide us
large amount of very interesting data. Based on Data Release 1 and 4
(DR1 and DR4, respectively) there were made two catalogues by Kleinman
et al. (2004) and Eisenstein et al. (2006), which contain 2551 and
9316 objects, respectively. Used identification methods did not allow
to find all white dwarfs with whole range of effective temperatures
and types in data sample. Kleinman et al. (2004) chose white dwarf
candidates based on dereddened colour indexes $u-g$ and $g-r$ and a
magnitude in $u$ filter. Selected white dwarf candidates was next
verified by visual inspection and prescribed its type - DA, DB, DZ
etc. After this procedure, spectra of selected objects were fitted by
theoretical DA and DB spectra.

Eisenstein et al. (2006) disposed larger amount of data so they used
more automatic selection procedure. They select stars fulfilled
applicable criteria (see details in Eisenstein et al. 2006). In this
manner they obtain sample of blue stars, which observational spectra
were fitted by theoretical spectra. Stars with parameters
characteristic for white dwarfs were next verified and classified
during different tests including visual inspection. This method has
some limitations, for example does not work for stars cooler than 8000K. 

We intend to find a method of the white dwarf candidates selection -
or, if it is possible, white dwarfs itself - based only on photometric
and spectral features, chosen to obtain the best distinguishing
between white dwarf stars and other type objects. Before we start
fitting theoretical model atmospheres to observing spectra.

Not for the first time we have used data from the Sloan Digital Sky
Survey (York et al. 2000). We chose Data Release 5 (DR5) and we
selected spectral data for 10\% randomly choosing objects. In this
manner we obtained very imposing amount of nearly 16000 different type
objects, including galaxies, quasars, stars, and of course white
dwarfs. Additionally we also used white dwarfs identified in SDSS DR4
by Eisenstein et al. (2006), treating these stars on the one hand as a
potential help to improve our white dwarfs selection method, on the
other hand as a control data.

\section{Balmer lines and their parameters}

The most characteristic feature of many DA white dwarf spectra is a
presence of wide, hydrogen absorption Balmer lines (except for the
hottest objects). It would appear that it should be a good selection
criteria of degenerated stars with high surface gravity, but not every
white dwarf have wide Balmer lines. Nevertheless we decided to check this
\begin{figure}[h]
\includegraphics[width=18.9pc]{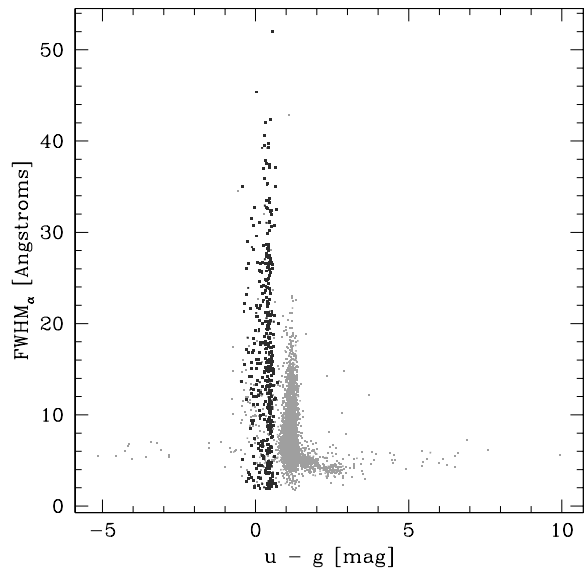}\hspace{1.5pc}%
\begin{minipage}[b]{16.5pc}\caption{\label{label}Colour index $u-g$ versus FWHM of the $H_{\alpha}$ Balmer line objects with $H_{\alpha}$, $H_{\beta}$, and $H_{\gamma}$ lines present in their spectra (dark gray dots). White dwarfs from SDSS DR4 belonging to our sample with detected Balmer lines was marked with black dots.
}
\end{minipage}
\end{figure}
possibility. Because of large amount (15998) of objects in our sample we
had to create proper software, which automatically finds hydrogen
lines from Balmer series, and then calculates its various parameters
like line widths at given depth, fluxes in lines etc. Fig. 1 shows one
of the obtained results, a dependency of FWHM  of Balmer $H_{\alpha}$
line on colour index $u-g$ for objects with $H_{\alpha}$, $H_{\beta}$, and
$H_{\gamma}$ lines present in their spectra. Clearly seeing bimodal
structure is not a by-product of our Balmer lines searching, but it is
a result of a bimodal distribution of the colour index $u-g$. A location
of the white dwarfs identified in SDSS DR4 (Kleinman et al. 2004), and
belonged to our sample with detected Balmer lines, denoted by black
dots, suggests, that all of the white dwarfs lies in the left part of
this diagram, with colour indexes $u-g < 0.7$. In general, this is not a
truth. Visual inspection shows, that some amount of white dwarf stars
also lies in the right part of the diagram, and have larger values of
colour indexes $u-g$. However, our requirement for the presence of
hydrogen Balmer lines in spectra reduces our sample to practically
star-like type objects only. In fact, it is decreases a number of
white dwarf candidates from 15998 to 4613. Unfortunately, reduced in
this way sample does not contain the hottest white dwarf stars,
because there is no Balmer lines visible in their spectra.

\section{Fluxes in given ranges} 

Colour indexes based on $ugriz$ photometry are not very useful for us,
generally because the filters are too wide. So we decide to define own
artificial filters, i.e. own wavelength ranges, in which we calculated
fluxes. Choosing our filters we intended to not contain any Balmer
\begin{figure}[h]
\begin{minipage}[t]{20pc}
\includegraphics[width=18.9pc]{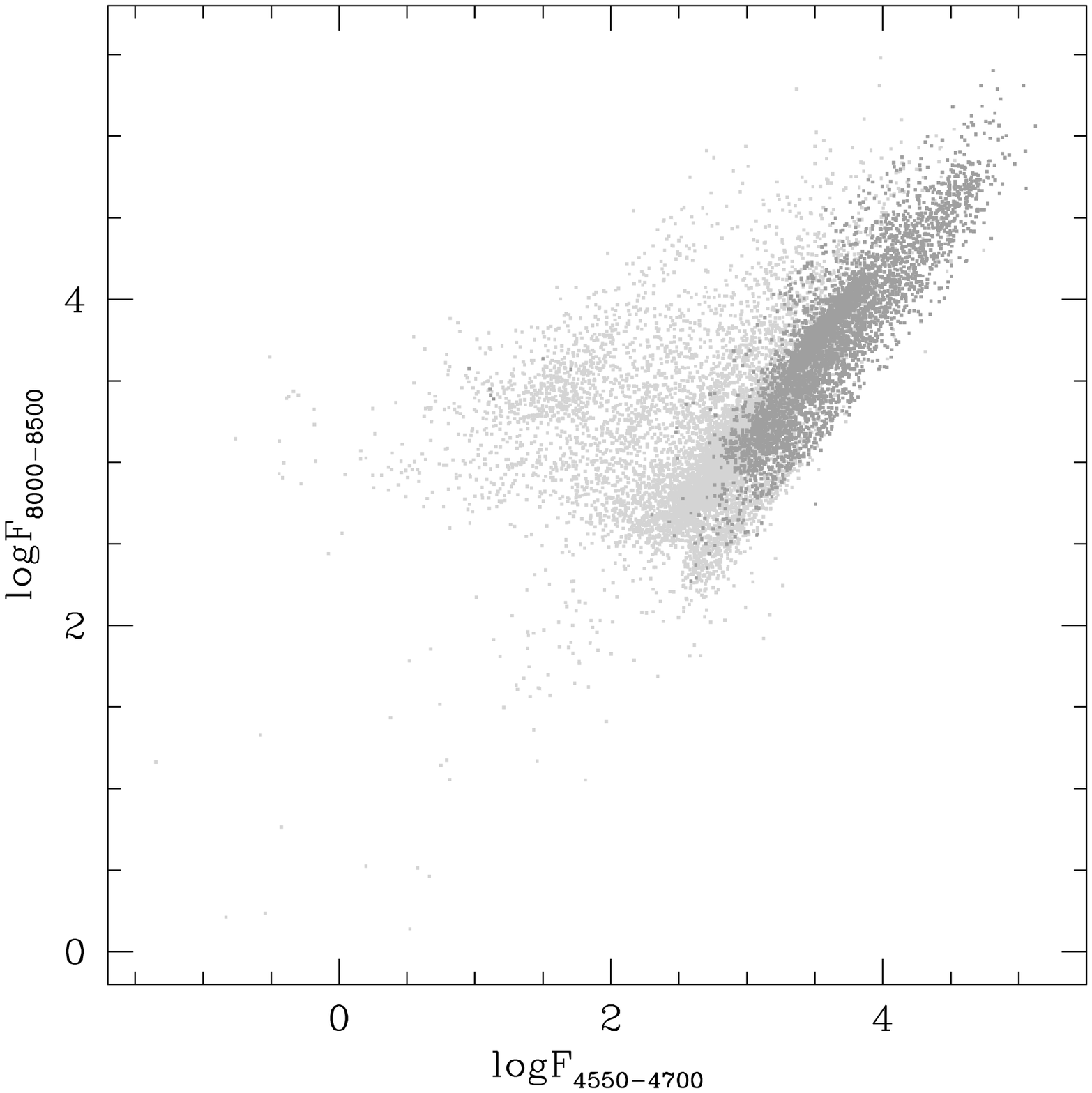}
\begin{minipage}{18pc}
\caption{\label{label}Logarithms of fluxes relation, calculated in two wavelength ranges: \mbox{$4550-4700${\AA}} and $8000-8500${\AA} for all of 15998 objects in our sample (light gray dots). Objects with detected $H_{\alpha}$, $H_{\beta}$, and $H_{\gamma}$ lines was marked by dark gray dots.} 
\end{minipage}
\end{minipage}
\begin{minipage}[t]{20pc}
\includegraphics[width=18.9pc]{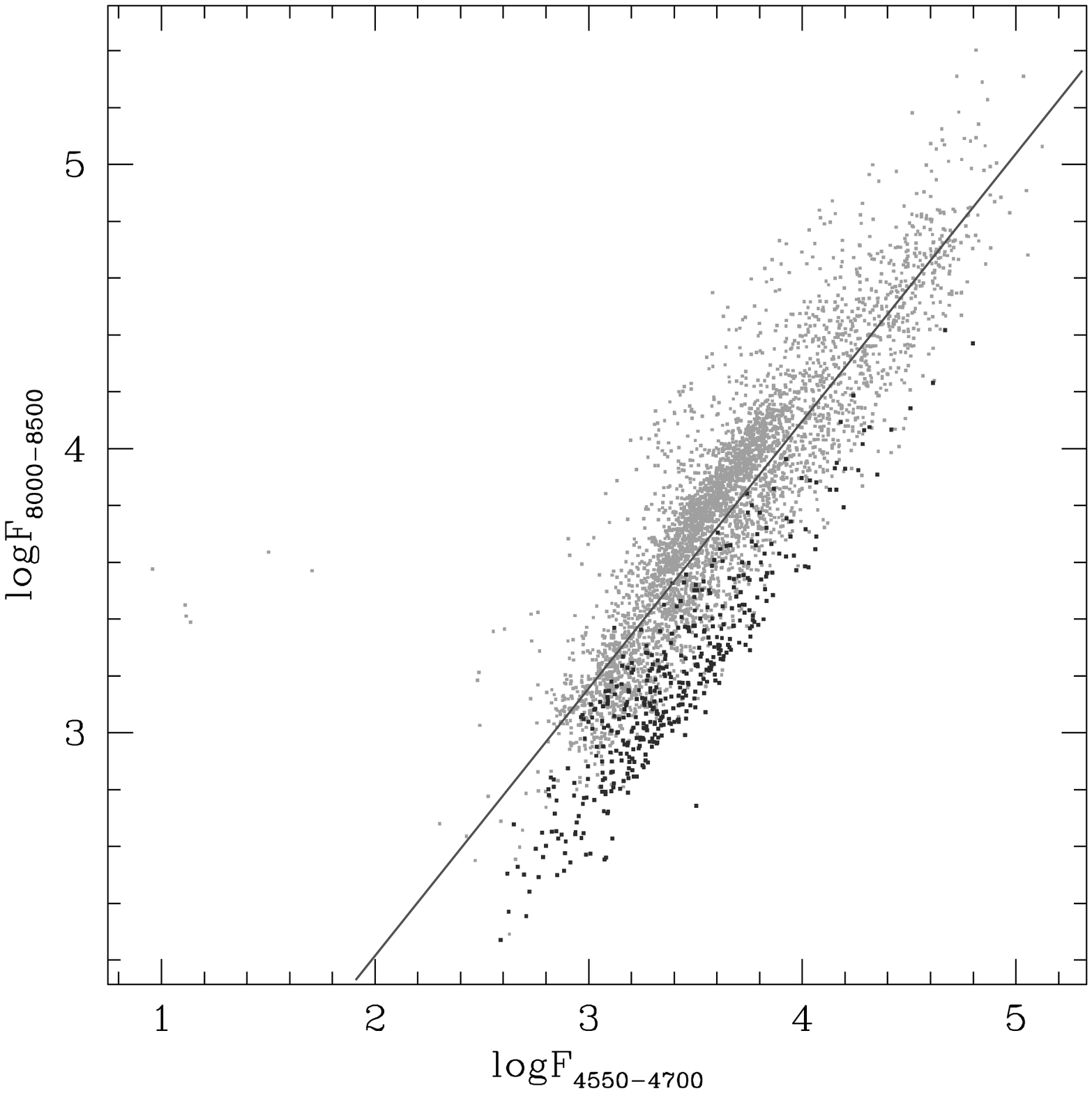}
\begin{minipage}{18pc}
\caption{\label{label}The same as in Fig. 2, but for 4613 objects with Balmer lines only (dark gray dots). Black dots denote white dwarfs from SDSS DR4. Linear function $y = 0.94118x + 0.33235$ is an approximate boundary between two regions visible in this diagram. Preliminary results of the visual inspection shows, that almost all of the 2202 objects located in the lower-right region looks like white dwarf stars.}
\end{minipage}
\end{minipage} 
\end{figure}
lines. In Fig. 2 we present a dependency of two flux logarithms,
calculated in two arbitrarily chosen wavelength ranges: $4550-4700${\AA} and
$8000-8500${\AA} for all of 15998 objects in our sample. Objects with
detected $H_{\alpha}$, $H_{\beta}$, and $H_{\gamma}$ lines (dark gray dots)
occupy well defined area, so we decided to watch them closer. Fig. 3
shows logarithms of fluxes in the same wavelength ranges as in Fig. 2,
but for 4613 objects with Balmer lines only. It is not very difficult
to see some structure in this diagram. At least part of the objects
divides to, not so bad, separate regions. This impression is
additionally magnified by black dots showing positions of the white
dwarfs from DR4, which occupy one of the mentioned above regions
only. We decided to approximate a boundary between these two areas by
the linear function $y = 0.94118x + 0.33235$, with empirically chosen
constants. Preliminary results of the visual inspection shows, that
spectra of almost all of the 2202 objects located in the lower-right
region looks like white dwarf spectra. Because of relatively poor
separation of these two areas, between 2411 objects belonging to the
upper-left region we still could find relatively small amount of white
dwarf stars, located close to the borderline.

\section{Conclusions}

Our method of selecting white dwarf candidates is based on searching for
objects with hydrogen Balmer lines visible in their spectra, and on
flux calculations in well selected wavelength ranges. Preliminary
results show that our method allows to select quite complete sample of
white dwarf stars, under the above assumptions. Of course this method
should be tested in detail, and it needs some improvements (for
example increasing separation between two areas in Fig. 3, probably by
choosing better wavelength ranges). It is quite possible, that using
this method we shall be able to find also the hottest white dwarfs, which
are too hot to show Balmer lines in their spectra, although we will
need the other criteria to select star-like type objects. We plan to
do this in future.

\subsection{Acknowledgments}

This work has been supported by the Polish Ministry of Science and
Higher Education grant No. N N203 4061 33. We also thank Institut
d'Estudis Espacials de Catalunya (IEEC) for financial support.

Funding for creation and distribution of the SDSS Archive has been
provided by the Alfred P. Sloan Foundation, the Participating
Institutions, the National Aeronautics and Space Administration, the
National Science Foundation, the US Department of Energy, the Japanese
Monbukagakusho, and Max Planck Society. The SDSS Web site is
http://www.sdss.org/ .

The SDSS is managed by the Astrophysical Research Consortium (ARC) for
the Participanting Institutions. The Participanting Institutions are
The University of Chicago, Femilab, the Institute for Advanced Study,
the Japan Participation Group, the Johns Hopkins University, Los
Alamos National Laboratory, the Max-Planck-Institute for Astronomy
(MPIA), the Max-Planck-Institute for Astrophysics (MPA), New Mexico
State University of Pittsburgh, Princeton University, the United
States Naval Observatory, and the University of Washington.

\section*{References}

\begin{thereferences}

\item Bergeron P, Saffer R A and Liebert J 1992  {\it ApJ} {\bf 477} 313

\item Eisenstein D J, Liebert J, Harris H C et al. 2006 {\it ApJS} {\bf 167} 40

\item Fleming T A et al. 1996 {\it A\&A} {\bf 316} 147

\item Green R F, Schmidt M and Liebert J 1986 {\it ApJS} {\bf 61} 305

\item Kleinman S J, Harris H C, Eisenstein D J et al. 2004 {\it ApJ} {\bf 607} 426

\item Liebert J and Bergeron P 1995 {\it White Dwarfs} eds D Koester and K Werner (Berlin: Springer) p12

\item Marsh M C, Barstow M A, Buckley D A et al. 1997a {\it MNRAS} {\bf 286} 369

\item Marsh M C, Barstow M A, Buckley D A et al. 1997b, {\it MNRAS} {\bf 287} 705

\item McCook G P and Sion E M 1987 {\it ApJS} {\bf 65} 603

\item Trumper J 1983 {\it Adv.Sp.Res.} {\bf 2} 241

\item Trumper J 1992 {\it QJRAS} {\bf 33} 165

\item York D G, Adelman J, Anderson J E Jr et al. 2000 {\it AJ} {\bf 120} 1579

\end{thereferences}

\end{document}